\documentclass[prl,twocolumn,amsmath,amssymb,citeautoscript]{revtex4}

\usepackage{amsmath}
\usepackage{amssymb}
\usepackage{epsfig}
\usepackage{graphicx}
\usepackage{wasysym}

\def \be{\begin{equation*}}
\def \ee{\end{equation*}}

\def \journal#1#2#3#4{{#1\ {\bf #2}, #3 (#4)}} 
\def \fig#1{{Fig.~\ref{#1}}}
\def \eqn#1{{Eqn.~(\ref{#1})}}

\def \vecS{\vec{S}}

\def \vecF{\vec{F}}

\def \vecu{\vec{u}}
\def \ising{\sigma}

\begin{document}
\title{Spin phonon induced colinear order and magnetization plateaus in triangular and kagome antiferromagnets. Applications to CuFeO$_2$.}
\author{Fa Wang$^{1,2}$
and Ashvin Vishwanath$^{1,2}$} \affiliation{ $^1$Department of
Physics, University of California, Berkeley, CA 94720
\\
$^2$Materials Sciences Division, Lawrence Berkeley National
Laboratory, Berkeley, CA 94720 }
\date{Printed \today}

\begin{abstract}
Coupling between spin and lattice degrees of freedom are important
in geometrically frustrated magnets where they can lead to
degeneracy lifting and novel orders. We show that moderate
spin-lattice couplings in triangular and Kagome antiferromagnets can
induce complex {\em colinear} magnetic orders. When classical
Heisenberg spins on the triangular lattice are coupled to Einstein
phonons, a rich variety of phases emerge, including the
experimentally observed four sublattice state and the five
sublattice 1/5th plateau state seen in the magneto-electric material
CuFeO$_2$. In addition we predict magnetization plateaus at 1/3,
3/7, 1/2, 3/5 and 5/7 at these couplings. Strong spin-lattice
couplings induce a striped colinear state, seen in
$\alpha$-NaFeO$_2$ and MnBr$_2$. On the Kagome lattice, moderate
spin-lattice couplings induce colinear order, but an extensive
degeneracy remains.
\end{abstract}
\maketitle


Frustrated magnets, in which classical ground states have multiple
accidental degeneracies, have been at the focus or renewed
attention. While a large body of theoretical work has focused on the
lifting of this degeneracy by thermal or quantum fluctuations
('order by disorder' \cite{villain})  in real materials, alternate
mechanisms might dominate. One mechanism that is always present is
the coupling of magnetism to the lattice (spin-phonon coupling).
Exploring the ramifications of such couplings is still in its
infancy and leads to new theoretical problems with consequences for
real materials. Finally, the subject is potentially important for
future applications - spin phonon couplings could give rise to
multiferroic systems, where magnetic order is coupled to
ferroelectricity \cite{mostovoy}.

The role of lattice distortions in promoting valence bond physics
(spin Pierels effect) is well known and consequences for spin 1/2
frustrated quantum magnets have been studied \cite{MilaJ1J2,SrCuBO}.
In the opposite large spin (semiclassical) limit, the role of
lattice distortions in selecting the {\em colinear} ground states
amongst the vast number of degenerate spin configurations of the
pyrochlore magnet has been emphasized\cite{Penc}. Further selection
of a unique colinear ground by various spin phonon interactions have
also been proposed \cite{tchernyshyov,Bergman}.

Here we will study the effect of spin-phonon interactions on the
classical ground states of the triangular and Kagome magnets. In
contrast to earlier applications, all ground states of the
antiferromagnets on these lattices are  non-colinear (120$^{\rm o}$)
configurations. Hence a finite strength of spin phonon coupling is
required to induce colinear states. We believe this to be the origin
of the puzzling colinear magnetism seen in several triangular
lattice materials, such as  CuFeO$_2$\cite{CuFeO},
$\alpha$-NaFeO$_2$\cite{McQueen}, and MnBr$_2$\cite{Sato}. In these
materials with $S=5/2$ filled shell moments, magnetic anisotropy is
likely to be (and in some cases known to be) very small, and cannot
be invoked to explain the colinear order. Furthermore, we study the
precise pattern of the colinear magnetic order induced within the
Einstein Site Phonon {\bf (ESP)} model \cite{Bergman}, which builds
in an energy cost for displacing a magnetic atom from its ideal
lattice position, and is parameterized by a single coupling
constant. Analytic arguments combined with numerical calculations
(simulated annealing) are used to establish the zero temperature
phase diagram of the classical spin models - which is briefly
summarized below. On the triangular lattice, increasing the
spin-phonon coupling leads to the phase diagram in Figure 1, where
the four sublattice zig-zag state ({\it Z} state) is the first
colinear state selected. At large couplings a four sublattice
parallel stripe state ({\it S} state) of up and down spins is
obtained. The phase diagram in a magnetic field is remarkably
complex, with magnetization plateaus at 1/5th, 1/3rd, 3/7th, 3/5th,
5/7th, and 1/2 occur in the parameter range where the {\it Z} state
appears, and also 1/9th and 7/15 at larger spin-phonon couplings. In
particular, the four sublattice {\it Z} state is typically converted
into a 1/5th plateau with a five sublattice spin structure on
increasing the field.

These are precisely the two colinear states observed in CuFeO$_2$,
\cite{CuFeO} an extensively studied triangular magnet, where the
complex four and five sublattice colinear structures described above
was seen. This was previously rather mysterious since the only
available models which captured such orders were Ising Hamiltonians
with large and very specific  second and third neighbor interactions
($J_2=0.45 J_1$, $J_3=0.75 J_1$)\cite{Mekata} (or more complicated
models, see e.g. Plumer\cite{Plumer}) which are unnatural since the
magnetic susceptibility is nearly isotropic in the paramagnetic
state. In contrast, our isotropic spin phon model involves a single
parameter - the spin phonon coupling. Predictions for higher
magnetization plateaus and other states for this material are made.
The magnetically induced electrical polarization observed in this
system is however not captured by our simple model, pointing to the
role of other interactions which we briefly discuss. The {\it S}
state is precisely the spin arrangement seen at low temperatures in
$\alpha-$NaFeO$_2$ and MnBr$_2$. Predictions for magnetic plateaus
in these materials are also made.

On the Kagome lattice too we find that beyond a critical coupling,
colinear ground states are obtained, but in contrast to the
triangular lattice, the manifold of these states has an extensive
entropy. Only at larger couplings is there a transition into a
unique ground state. This may be of relevance to the recently
studied Kagome staircase compound, Mn$_3$V$_2$O$_8$\cite{Cava}. If
the spin phonon mechanism is as widely relevant as some of our
results suggest, theoretical studies will have to move beyond their
current focus on purely rigid lattice models.




{\bf Spin Phonon Model:} The minimal spin phonon coupling arises
from the dependence of the exchange coupling on separation between
the magnetic ions $\bar{J}(r)$. If sites $i$ and $j$ approach each
other by $u_{ij}$, then
 the spin phonon Hamiltonian then is:
\begin{equation} H=\bar{J}\sum_{<ij>}{(1-\alpha_{ij}u_{ij})\vecS_i\cdot\vecS_j}
+H_{\rm lattice}(\{\vecu_i\}) \end{equation} where $\alpha =
\bar{J}^{-1}\partial \bar{J}/\partial r$, $H_{\rm lattice}$ is the
phonon Hamiltonian and $\vecu_i$ is the displacement of site $i$.
Hence, $u_{ij}=(\vecu_i-\vecu_j)\cdot \hat{e}_{ij}$ is the relative
displacement projected on the bond $ij$ ($\hat{e}_{ij}$ is the unit
vector from site $i$ to $j$).
Two types of phonon Hamiltonians have been proposed. First, the bond
phonon model of Penc et al.\cite{Penc} where $H_{\rm
lattice}=(K/2)\sum_{<ij>}{u_{ij}^2}$. This model assumes that bond
displacements are independent, which may be an oversimplification
for many lattices. Integrating out the phonon degree of freedom here
generates just the biquadratic term $-b
\bar{J}\sum_{<ij>}{(\vecS_i\cdot\vecS_j)^2}$. (The phonon dynamics
can be neglected if its frequency is much larger than magnetic
energy scales). While for sufficiently
 large $b$, colinear states result, the model is still
 highly degenerate (Ising antiferromagnet ground states on the triangular or Kagome
 magnets). The selection by quantum fluctuations will be discussed
 elsewhere \cite{bondphonon}. Instead, here we turn to the second model, the ESP model of Bergman et al.\cite{Bergman}
\begin{equation}
H_{\rm lattice}=(K/2)\sum_{i}{\vecu_i^2}
\end{equation}
which in contrast to the bond phonon model, respects the inevitable
correlations between bond length arising from the underlying lattice
structure.
The ESP model describes a dispersionless optical phonon branch. More
realistic phonon model informed by the detailed crystal structure of
the material and acoustic phonons could generate longer-range
effective spin interactions that are hard to deal with. In the
interests of simplicity and generality we will restrict attention to
the ESP model. Integrating out the lattice displacement $\vecu_i$
results in the effective spin Hamiltonian:
\begin{equation}
H_{ESP}=\bar{J}\left [ \sum_{<ij>}{\vecS_i\cdot\vecS_j}-\bar{c}
\sum_{i}{\vecF_i^2} \right ]\label{equ:sitephononmodel}
\end{equation}
where $\bar{c}=\alpha^2 \bar{J}/(2K)$ is a positive dimensionless
coupling and we have defined the dimensionless `force' on site $i$
as $\vecF_i=\sum_{j\in\ {\rm neighbors\ of\
}i}(\vecS_i\cdot\vecS_j)\hat{e}_{ij}$, the sum in this definition is
over the (six) neighbors of site $i$. The spin-phonon interaction
seeks to maximize the force $\vec{F}$, which will result in gaining
the maximum spin interaction energy.  Note, the second term in Eqn.
\ref{equ:sitephononmodel} generates interactions involving three
adjacent spins
$\hat{e}_{ij}\cdot\hat{e}_{jk}(\vecS_i\cdot\vecS_j)(\vecS_j\cdot\vecS_k)$.

 {\bf
Triangular Lattice:} In the following we will consider a single
triangular lattice sheet governed by the Hamiltonian in
\eqn{equ:sitephononmodel} i.e. with nearest neighbor
antiferromagnetic interactions and additional interactions generated
by the spin-phonon term.
We focus on the zero temperature phase diagram, as a function of the
single parameter $c$ and subsequently in an applied magnetic field.
Since we are primarily interested in large spin (eg. $S=5/2$), we
focus on classical spins, where we can write $\vec{S}_i=S\hat{n}_i$,
where $\hat{n}_i$ is a unit vector. This leads to an extended
classical Heisenberg model of unit vectors with rescaled couplings
$J=S^2\bar{J}$ and $c=S^2\bar{c}$ . If we choose to
restrict to Ising states, $\vecS_i=\ising_i\vecS$, this effective
spin Hamiltonian simplifies to an Ising model with nearest-, second-
and third-neighbor coupling, $J-c J$, $c J$ and $c J$ respectively.
However, in this model the second and third neighbor couplings are
constrained to be strictly equal. This might be a rationalization
for the large second- and third-neighbor couplings used in previous
Ising models \cite{Mekata}.

\begin{figure}
\includegraphics{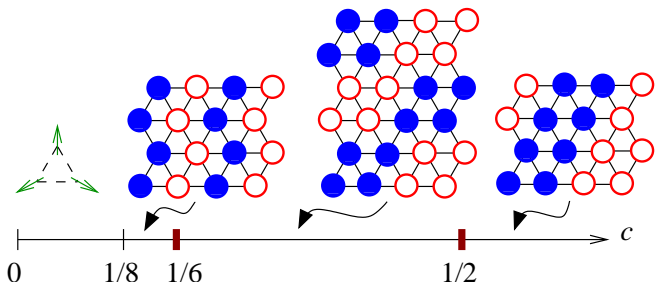}
\caption{(Color online) zero field phase diagram of the triangular
lattice ESP model. On increasing the spin phonon coupling $c$, the
120$^o$ state is followed first by the zig-zag {\it Z} state, then
an 8 sublattice state  and finally the stripe {\it S} state.
Accidental degeneracies (thick red ticks) only occur at the phase
boundaries $c=1/6,\,1/2$. Solid (hollow) blue (red) circles are up
(down) spins.} \label{fig:zerohphsaediagram}
\end{figure}

Consider the ground state for classical spins on raising $c$. While
at $c=0$ the the regular 120$^0$ pattern of O(3) spins on the
triangular lattice is realized, this is expected to survive to
finite $c$ as well. The ground state energy per site is
$E_0/J=-3/2$, and since the force vanishes in this state, it is
independent of $c$. While a full numerical solution is required (and
provided below) for the phase diagram of this model, we can make
some plausible analytic arguments which will be confirmed by the
numerics. Clearly, colinear states are preferred for large $c$ since
they give rise to the maximum force. However, near the phase
boundary with the 120$^0$ states, the exchange $J$ will presumably
be important, and hence we restrict attention to those colinear
states that best satisfy $J$. These are nothing but the ground
states of the triangular lattice Ising antiferromagnet (TLIA) with
nearest neighbor exchange, with two (one) up and one (two) down
spins per triangle. We can then ask, which configuration within this
manifold optimizes the force term? This question can be rigorously
answered - it is the {\it Z} state. Using the dimer representation
of the TLIA states, where a dimer is drawn orthogonal to each
unsatisfied bond, and leads to a hard core dimer configuration on
the honeycomb lattice, we see that the `force' on site $i$ is
determined by the dimer configurations on the hexagon surrounding
site $i$. By checking the possible dimer configurations on a
hexagon,
 we see that the force $|\vecF_i|=2$ if there is one dimer in the
hexagon {\em or} two not-opposite dimers; otherwise $|\vecF_i|=0$.
The ground state maximizes $\sum_{i}{\vecF_i^2}$. Hence it must have
two dimers in every hexagon (since on average there are two dimers
per hexagon and a configuration with a one dimer hexagon implies
also a hexagon with three dimers, which experiences no force).
Combining this with the condition that the two dimers cannot be on
opposite sides leads us to the unique zigzag state {\it Z}, with a
four sublattice unit cell as shown in Fig.
\ref{fig:zerohphsaediagram}, and ground state energy per site
$E_0/J=-1-4c$.

The full phase diagram is obtained using simulated annealing on
lattices with periodic boundary condition and various sizes up to
$10\times 10$, and choosing the state with lowest energy per site.
Simulations on each size were done by an exponential annealing
schedule from $\beta J=0.1$ with a random initial state to $\beta
J=1000$, with a total of 20000$\sim$40000 sweeps, the whole process
was repeated 10 times to ensure stability of results. Site-update
with Metropolis dynamics was used. While the algorithm does not
guarantee convergence to the ground state we nevertheless believe an
accurate picture emerges since all analytic expectations have been
met, and we have not been able to guess ground states with better
energies.

\begin{figure}
\includegraphics{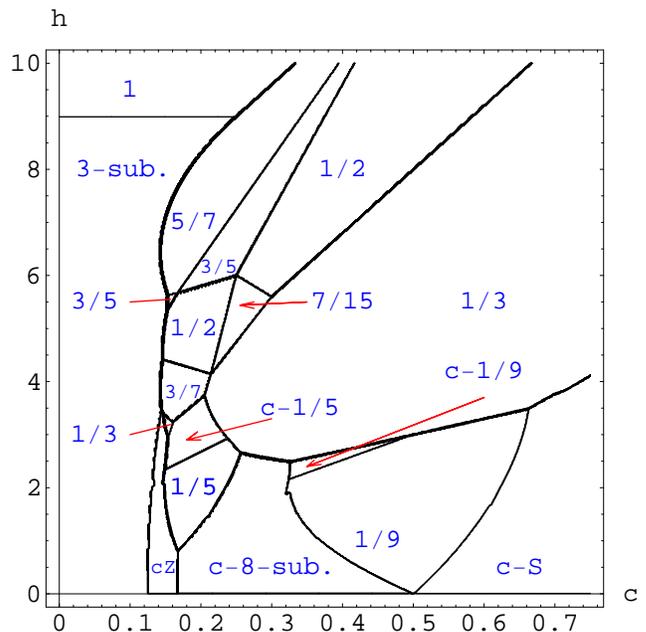}
\caption{$T=0$ phase diagram of the triangular lattice ESP model.
The vertical axis is the normalized field $h=H/J$. The fraction $f$
label magnetization plateaus, while c$-f$ are canted (non-colinear)
states deriving from them. Also, cZ is the canted {\it Z} state
while the $3-sub$ states at small $c$ also occur in the pure
Heisenberg model in a field.} \label{fig:annealingphasediagram}
\end{figure}

The numerically obtained phase diagram in zero field is shown in
Fig. \ref{fig:zerohphsaediagram}. At $c=1/8$ the 120$^o$ state is
replaced by the four sublattice {\it Z} state, with alternating
zig-zags of up and down spins and ground state energy per spin
$E_0/J=-1-4c$. This is stable till $c=1/6$, where an 8 sublattice
state ($E_0/J=-10c$) takes over. Beyond $c=1/2$, the {\it S} state
of up-up-down-down stripes ($E_0/J=1-12c$) is found and persists to
large couplings. At the transition points c=1/6 and c=1/8, there are
additional accidental degeneracies, and quantum effects could be
important in resolving these\cite{unpublished}.


The phase diagram in a magnetic field $H_h= -J\sum_ih n^z_i$ is
remarkably rich. For small $c$ where the 120$^0$ state is realized,
a highly degenerate set of states \cite{chubukov} well known for the
Heisenberg triangular antiferromagnet are obtained. Since they all
have vanishing force contributions, the spin phonon interaction does
not split this degeneracy. At larger values of $c$ , the simulation
shows a plethora of plateau states, which we briefly discuss here
and leave details to \cite{unpublished}. Interestingly the
1/5-plateau with the pattern observed in CuFeO$_2$ occur for a wide
range of parameter $c$. For the parameter interval $0.14<c<1/6$ our
model shows both the zigzag {\it Z} ground state at zero field and
the 1/5-plateau in magnetic field, as in CuFeO$_2$. Other prominent
plateaus that occur in the range of $c$ where the {\it Z} state
appears are the 3/7th and 5/7th states with 7 site unit cell, a 1/2
magnetization plateau with an 8 site unit cell and two distinct
3/5th plateaus with 5 sites per unit cell. There is also a small
region of 1/3 plateau, with a 12 site unit cell. The evolution of a
plateaus state with increasing field can proceed in two ways. Either
it can jump directly to another plateau, or undergo a canting
transition, where the field direction is not parallel to the
staggered moment. For example, the zero field {\it Z} state in the
fully isotropic model undergoes a spin-flop transition immediately
on applying a field, and the staggered moment is orthogonal to the
field and the induced uniform magnetization. The evolution of the
1/5 plateau state on increasing the field however is continuous,
with a gradual tilting of the staggered component away from the
field. This phase boundary can be calculated analytically and agrees
very well with the simulations. Such canted states are labeled $c-f$
in the figure (where $f$ refers to the plateau they derive from).
Such states are of course absent in Ising model studies
\cite{Mekata}. Other plateaus occur for larger $c$, which will be
discussed in detail elsewhere \cite{unpublished}. Here we note that
the 1/9th plateau that extends all the way down to zero field occurs
because of the accidental degeneracy at the point $c=1/2$ which
includes finite magnetization configurations, with a maximum
magnetization of 1/9th, which is selected by the field. Amusingly,
the most obvious 1/3 plateau expected for a triangular lattice,
consisting of up,up and down spins on the three sublattices, does
not occur (the 1/3 plateau at high fields involves a 12 spins).

\begin{figure}
\includegraphics{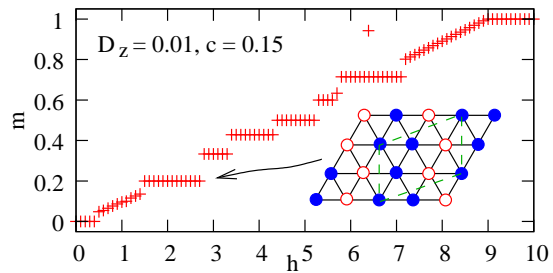}
\caption{(Color online) Predicted magnetization ($m$) curve for
CuFeO$_2$ with $c=0.15$ and 1\% anisotropy. Plateaus occur at
$m=0,\,1/5,\,1/3,\,3/7,\,1/2,\,3/5,\,5/7$ and of course at $m=1$.
One unit of field is $\sim 10$ Tesla. Inset: the $m=1/5$ state.}
\label{fig:plateausCuFeO2}
\end{figure}

{\bf CuFeO$_2$ and other materials:} In CuFeO$_2$ the 4 sublattice
{\it Z} state is observed, which persists in a field upto $B<6$
Tesla. At higher fields $B>14$Tesla, the 5 sublattice 1/5th
magnetization plateau is observed. We note that both these states
occur in our spin phonon model when $0.14<c<1/6$. To estimate the
spin phonon coupling $c=\alpha^2J/2K$ in CuFeO$_2$, we use
$J=39$Kelvin (from the measured Weiss constant \cite{Patrenko}) and
estimate $\alpha \sim 7$ and $K \sim 10,000$Kelvin \cite{SrCuBO},
which gives $c\sim 0.1$ which is in the right ball park. While an
isotropic spin model with magnetic order cannot have a magnetization
plateau centered at zero field, even a very small magnetic
anisotropy (e.g. an easy axis anisotropy $-D\sum_i S_{zi}^2$) can
produce the observed zero magnetization plateau, since the plateau
width $\Delta B$ is readily seen to be $\Delta B \propto \sqrt{DJ}$.
A 1\% anisotropy $D/J$ produces the right plateau width
\cite{Patrenko}. The magnetization profile as a function of field at
$c=0.15$ with a 1\% easy axis anisotropy is shown in Fig.
\ref{fig:plateausCuFeO2} (the field scale $J/g\mu_BS$ is
$\sim$10Tesla per unit from the estimated value of $J$). The higher
field magnetization plateaus and structures are predicted for future
experiments on single crystals. Existing pulsed field measurements
on powder samples reveal a sequence of anomalies at different fields
up to full polarization, but the magnetization plateaus and
structures remain to be conclusively identified\cite{Ajiro}. Lastly
we note that the ferroelectric phase with incommensurate spiral
order observed experimentally in CuFeO$_2$ in the field range
$7<B<14$Tesla\cite{Ramirez} is not produced here, indicating the
importance of other lattice couplings eg. to the oxygen atoms
mediating the superexchange interaction. The up up down down stripe
pattern for $1/2<c$ has been observed as the ground state for some
materials with triangular lattice structure
 such as $\alpha$-NaFeO$_2$,\cite{McQueen} and MnBr$_2$,
 \cite{Sato}. Finding the predicted 1/9th and 1/3 plateaus in these
 materials would be a check of the spin phonon origin of these
 states.

{\bf Kagome Lattice:} The Einstein site phonon model on the Kagome
is virtually identical to the triangular case, except that the lower
symmetry in this case (lack of site centered 60$^{\rm o}$ rotation
symmetry) allows for anisotropic confining potential on the atoms.
For simplicity, we assume an isotropic confining potential, but the
main results are independent of this assumption.

\begin{figure}
\includegraphics{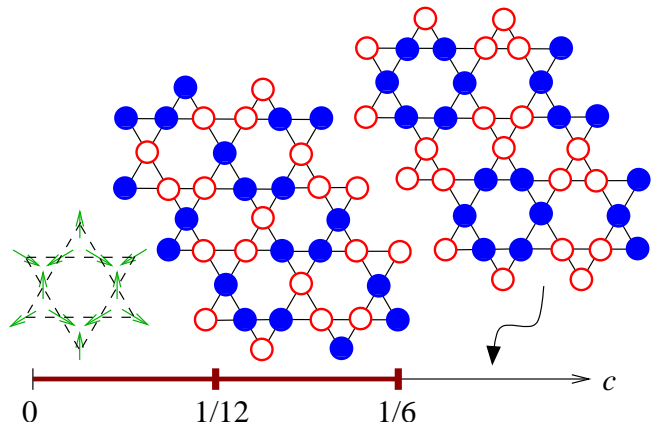}
\caption{(Color online) zero field phase diagram of the ESP model on
kagome lattice. Extensive degeneracy (marked red) persists into
$1/12<c<1/6$ where colinear states occur - representative
configurations are shown.} \label{fig:zerohphsaediagram-kagome}
\end{figure}

Simulated annealing was applied to this model with similar settings
as the triangular case. The zero-field phase diagram is presented in
\fig{fig:zerohphsaediagram-kagome}. For small $c$ we still get the
ground states of the pure Heisenberg model, which are known to be
extensively degenerate. For $c>1/12$ we get colinear states, but in
the range $1/12<c<1/6$ an extensive degeneracy remains
\cite{sen_damle_vishwanath}. Even more interestingly the zero-field
ground states can have arbitrary magnetization ranging from $-1/9$
to $1/9$ per site. Therefore, in this zero temperature classical
model, applying a small field will immediately induce a
$1/9$-magnetization-plateau state. We expect that thermal and/or
quantum fluctuation can lift this accidental degeneracy which is
left for future work \cite{unpublished}. Further increasing $c$
beyond $1/6$ pushes the system into a unique colinear states (see
Fig. \ref{fig:zerohphsaediagram-kagome}).

We have shown that spin lattice couplings can induce a rich variety
of complex colinear orders on the triangular and Kagome lattices.
The moderate spin-phonon couplings strengths required make this a
viable mechanism to explain similar ordering patterns seen in
materials like CuFeO$_2$.  Our results underline the need to go
beyond the current focus on purely rigid lattice spin models. Future
work will study the effect of quantum fluctuations and valence bond
states that naturally arise from disordering these colinear
configurations. Including other atomic displacements might account
for interesting magnetoelectric phenomena, and guide the search for
multiferroic materials. We acknowledge support from LBNL DOE-504108
and useful discussions with K. Damle and R. Cava.


\end{document}